\documentclass[]{mn2e}
\usepackage{epsfig}
\def\lsim{\vcenter{\hbox{$<$}\offinterlineskip\hbox{$\sim$}}}

\title[Mass loss and the ISM in globular clusters]
{Stellar mass loss and the Intra-Cluster Medium in Galactic globular clusters:
a deep radio survey for H\,{\sc i} and OH}
\author[Jacco Th. van Loon et al.]{Jacco Th. van Loon$^{1}$\thanks{E-mail:
jacco@astro.keele.ac.uk}, Sne\v{z}ana Stanimirovi\'{c}$^{2}$, A. Evans$^{1}$
and Erik Muller$^{3,4}$\\
$^{1}$Astrophysics Group, School of Physical \& Geographical Sciences, Keele
      University, Staffordshire ST5 5BG, United Kingdom\\
$^{2}$Radio Astronomy Lab, University of California at Berkeley, 601 Campbell
      Hall, Berkeley CA 94720, USA\\
$^{3}$Arecibo Observatory, National Astronomy and Ionosphere Center, HC3 Box
      53995, Arecibo PR 00612, USA\\
$^{4}$CSIRO Australia Telescope National Facility, PO Box 76, Epping NSW 1710,
      Australia}
\date{Submitted 2005}
\pagerange{\pageref{firstpage}--\pageref{lastpage}}
\pubyear{2005}
\begin{document}
\maketitle
\label{firstpage}
\begin{abstract}
We present the results of a survey, the deepest to date, for H\,{\sc i}
emission at 21 cm and OH emission at 18 cm (lines at 1612, 1665, 1667 and
1720 MHz) in the direction towards the Galactic globular clusters M\,15, M\,2,
NGC\,6934, NGC\,7006 and Pal\,13. The aim is to measure the amount of hydrogen
in the intra-cluster medium (ICM), and to find OH masers in the circumstellar
envelopes of globular cluster red giants. We present a tentative detection of
0.3 M$_\odot$ of neutral hydrogen in M\,15 and possible detections of neutral
hydrogen in M\,2 and Pal\,13. We derive upper limits to the neutral hydrogen
content of NGC\,6934 and NGC\,7006. No OH emission is detected. We also
present deep H\,{\sc i} data of the northern tip of the Magellanic Stream
behind Pal\,13.
\end{abstract}

\section{Introduction}

Globular clusters (GCs) contain red giants, both first ascent Red Giant Branch
(RGB) stars and Asymptotic Giant Branch (AGB) stars. These red giants have
evolved from 0.8--1 M$_\odot$ main-sequence stars, and are believed to lose
10--20 per cent of their mass on the RGB and a similar amount on the AGB
before ending as low-mass white dwarfs. This mass loss is induced by radial
stellar pulsation, and further driven by radiation pressure on dust grains
that form in the cool, dense atmosphere. The GC's Intra-Cluster Medium (ICM)
is thus continuously replenished with neutral material. The wind outflow
velocities are with $v_{\rm wind}\sim$10 km s$^{-1}$ smaller than the escape
velocity $v_{\rm esc}\sim$20--50 km s$^{-1}$ of a typical GC. Hence of order
100 M$_\odot$ of interstellar material must have accumulated within the GC
(Tayler \& Wood 1975) before it is removed by ram pressure upon crossing the
Galactic plane, which typically happens once every $\sim$$10^8$ yr
(Odenkirchen et al.\ 1997).

Circumstellar dust has been found around red giants in the GCs M\,15, M\,54,
NGC\,362, NGC\,6388, $\omega$\,Cen and 47\,Tuc (Ramdani \& Jorissen 2001;
Origlia et al.\ 2002). Origlia et al.\ (1997) possibly detected CO emission
from Mira variables in $\omega$\,Cen. Further evidence for stellar mass loss
in GC red giants includes blue-asymmetric absorption line profiles in optical
spectra (Bates, Kemp \& Montgomery 1993), H$\alpha$ line emission (Cohen 1976)
which however is complicated by the presence of a chromosphere (Dupree,
Hartmann \& Smith 1990), and stellar pulsation and associated levitation of
the stellar atmosphere (Frogel, Persson \& Cohen 1981; Frogel \& Elias 1988).
In addition, 4 Planetary Nebulae (PNe) have been found in as many GCs (Jacoby
et al.\ 1997).

However, the ICM has been more elusive.

{\bf Dust:} Dust comprises at most one per cent of the mass, but it is easily
detected. IR observations (IRAS, ISO) placed upper limits of $M_{\rm
dust}\sim$10$^{-3}$ M$_\odot$ (Lynch \& Rossano 1990; Knapp, Gunn \& Connolly
1995; Origlia, Ferraro \& Fusi Pecci 1996; Hopwood et al.\ 1999). Observations
at mm wavelengths yield similar limits (Penny, Evans \& Odenkirchen 1997),
with a possible detection of $M_{\rm dust}\sim$10$^{-2}$ M$_\odot$ in the
metal-rich GC NGC\,6356 (Hopwood et al.\ 1998). The only secure detection,
$M_{\rm dust}\sim5\times$10$^{-4}$ M$_\odot$ of cold dust was found in the
metal-poor GC M\,15 (Evans et al.\ 2003).

{\bf Molecular gas:} CO is a good tracer of molecular gas, which is mostly in
the form of H$_2$, but attempts to detect CO yielded upper limits of $M_{\rm
gas}\sim$0.1 M$_\odot$ (Smith, Woodsworth \& Hesser 1995; Leon \& Combes 1996;
Hopwood 2001), with a possible detection at a similar level in 47\,Tuc
(Origlia et al.\ 1997). Searches for OH 1612 MHz maser emission (Knapp \& Kerr
1973; Frail \& Beasley 1994) only produced foreground sources, and no H$_2$O
masers have been found (Cohen \& Malkan 1979). The most sensitive OH search to
date, in the 1665+7 MHz main lines at Arecibo by Dickey \& Malkan (1980),
resulted in no detection.

{\bf Atomic gas:} Faulkner et al.\ (1991) detected neutral hydrogen (H\,{\sc
i}) at 21 cm in NGC\,2808 and inferred the presence of $M_{\rm gas}\sim$200
M$_\odot$, but other attempts to detect H\,{\sc i} have so far resulted in
upper limits of only a few M$_\odot$ (e.g., Knapp, Rose \& Kerr 1973; Smith et
al.\ 1990) --- down to as little as $\sim$0.3 M$_\odot$ in the most sensitive
searches performed at Arecibo (Birkinshaw, Ho \& Baud 1983; H\,{\sc i} detected
near M\,56 is probably associated with intervening material).

{\bf Ionized gas:} Hot stars such as blue horizontal branch stars or post-AGB
stars produce a radiation field with $T_{\rm rad}>10,000$ K, which might
ionize the ICM. Searches for free-free continuum or H$\alpha$ line emission
have set limits of $M_{\rm gas}\sim$1 M$_\odot$ (Knapp et al.\ 1996: VLA, 8.4
GHz; Faulkner \& Freeman 1977; Smith, Hesser \& Shawl 1976; Hesser \& Shawl
1977; Grindlay \& Liller 1977). On the other hand, the presence of a
population of free electrons in 47\,Tuc is inferred from pulsar timing
observations (Freire et al.\ 2001).

We have performed a new, sensitive search at Arecibo in the 21 cm H\,{\sc i}
transition and, in parallel, in all four 18 cm OH transitions, to investigate
the atomic and molecular content of the ICM and to search for circumstellar OH
masers within a small selection of Galactic GCs.

\section{Observations}

\subsection{Radio L-band spectroscopy}

The Arecibo radio telescope\footnote{The Arecibo Observatory is part of the
National Astronomy and Ionosphere Center, operated by Cornell University under
a cooperative agreement with the National Science Foundation.} was used
between 17 and 23 September 2004 with the L-wide receiver to measure the
H\,{\sc i} line at 1420 MHz, the OH mainlines at 1665 and 1667 MHz and the OH
satellite lines at 1612 and 1720 MHz in the direction of a selection of
Galactic GCs (Table 1).

Each of the four boards of the correlator was centered on one of the lines'
laboratory frequency, with board 3 being centered on 1666.38 MHz in between
the two OH main lines. The correlator was set for 9-level interleave sampling
at two orthogonal polarizations, using a bandwidth of 6.25 MHz for the H\,{\sc
i} and 3.125 MHz for the OH lines. The four spectra, obtained with 1-sec
integration dumps, were sampled with 1024 channels, resulting in a channel
width of 1.29 and 0.57--0.53 km s$^{-1}$ for the H\,{\sc i} and OH\,1612--1720
MHz lines, respectively.

The system temperature was typically 26--28 K at 1420 MHz, at a gain of
$\sim$10 K Jy$^{-1}$. Calibration from correlator counts to the antenna
temperature units was performed with respect to the noise diode (``hcal'') of
known strength. The system was validated at the start of the first night
through the observation of the bright circumstellar OH maser emission from the
Galactic star RX Sge.

We followed a standard ON/OFF position switching strategy, pointing
alternately at the centre of the target and at a ``blank sky'' position, while
tracking exactly the same azimuth and zenith angle range for both ON and OFF
scans. Typical integration time per scan was 5 min, followed by an ON/OFF
observation of the noise diode for 10 sec. The total accumulated on-target
integration times are listed in Table 2. As an additional experiment, to
better sample the emission from intervening matter, we also performed sparse
mapping of M\,15, M\,2 and Pal\,13 by obtaining ON spectra at 6--8 positions
around the target's centre separated by 3.5 arcmin. Note that the FWHM for the
L-wide receiver is 3.4 arcmin at a frequency of 1420 MHz, and 2.9 arcmin at a
frequency of 1666 MHz (Heiles et al.\ 2001, with updates on
http://www.naic.edu/$\sim$phil/sysperf/sysperf.html).

The spectra were processed in IDL, using standard Arecibo routines to
calibrate the individual ON and OFF spectra (http://www.naic.edu/$\sim$phil/).
The combined OFF spectrum was smoothed by 7 channels (H\,{\sc i}) or 15
channels (OH), respectively, to provide a reference which is practically noise
free but which still contains all of the slow baseline structure. The
baseline-corrected spectra were then derived by taking the difference between
the combined ON spectrum and this reference.

\subsection{Target selection}

The targets were selected from the Harris (1996: revision 22 June 1999)
catalogue of 147 Galactic GCs, with $-1 <$ Declination $< +36$ deg and
Galactic latitude $|b_{\rm II}| > 18$ deg. Some of their properties are
summarised in Table 1. Both M\,15 and M\,2 are massive GCs with large tidal
radii, and any ICM may therefore be distributed over many arcminutes. However,
most of the stellar mass is well confined within the Arecibo beam of
$\sim3^\prime$, with M\,15 having experienced core collapse, and any ICM must
also be centrally concentrated to some degree. The GCs NGC\,6934 and NGC\,7006
have large velocities with respect to the Galactic disk, minimising the risk
of contamination with foreground emission.

M\,15 is given the highest priority since circumstellar dust (Origlia et al.\
2002) and intra-cluster dust (Evans et al.\ 2003) have been detected in it, as
well as the PN Ps\,1 (Alves, Bond \& Livio 2000 and references therein). The
high central velocity dispersion of $\sim10$ to 15 km s$^{-1}$ is interpreted
by some as evidence for the presence of an intermediate-mass black hole
(Gerssen et al.\ 2002), although recent N-body models do not require such a
black hole (McNamara, Harrison \& Anderson 2003).

It has been argued that the small GC Pal\,13 is in the process of being
tidally disrupted unless it is dark-matter dominated (Siegel et al.\ 2001;
C\^{o}t\'{e} et al.\ 2002). Blecha et al.\ (2004) measure a systemic velocity
of $v_{\rm LSR}=32.2\pm0.4$ km s$^{-1}$ and a radial velocity dispersion of
$\sigma_{\rm v}=0.9\pm0.3$ km s$^{-1}$. C\^{o}t\'{e} et al.\ show evidence for
the velocity dispersion to rise from $<1$ km s$^{-1}$ in the cluster core to
$\sim3$ km s$^{-1}$ (i.e.\ comparable to the escape velocity) at a projected
distance of $1.4^\prime$ from the core.

%
%
\begin{table*}
\caption[]{The observed GCs, with parameters from Harris (1996: revision 22
June 1999) unless otherwise indicated. Listed are: name, equatorial (J2000)
and Galactic coordinates, distance $d_\odot$ to the Sun, reddening $E_{\rm
B-V}$, systemic velocity $v_{\rm LSR}$ with respect to the Local Standard of
Rest, central radial velocity dispersion $\sigma_{\rm rad}$, central escape
velocity $v_{\rm esc}$ (from Webbink 1985), metallicity [Fe/H] with respect to
solar, and the core radius $r_{\rm core}$, half-mass radius $r_{\rm M/2}$ and
tidal radius $r_{\rm tidal}$. References are as follows: a=Blecha et al.\
(2004); b=C\^{o}t\'{e} et al.\ (2002); c=Gerssen et al.\ (2002); d=McNamara et
al.\ (2003); e=Pryor et al.\ (1986); f=Siegel et al.\ (2001); g=Webbink
(1985).}
\begin{tabular}{lcccccclccrcccc}
\hline\hline
Object                 &
RA$_{\, 2000}$         &
\llap{D}ec$_{\, 2000}$ &
l$_{\rm II}$           &
b$_{\rm II}$           &
$d_\odot$              &
\llap{$E$}$_{\rm B-V}$ &
                       &
$v_{\rm LSR}$          &
$\sigma_{\rm rad}$     &
$v_{\rm esc}$          &
[Fe/H]                 &
$r_{\rm core}$         &
$r_{\rm M/2}$          &
$r_{\rm tidal}$        \\
                       &
(h\ m\ s)              &
($\circ\ \prime\ \prime\prime$) &
($\circ$)              &
($\circ$)              &
(kpc)                  &
\llap{(}mag\rlap{)}    &
                       &
\multicolumn{3}{c}{\llap{-----}------ (km s$^{-1}$) ---------} &
                       &
\multicolumn{3}{c}{------ (arcmin) ------} \\
\hline
NGC\,6934               &
20 34 11.6              &
7 24 15                 &
52.10                   &
$-$18.89                &
17.4                    &
0.09                    &
                        &
\llap{$-39$}$6.7\pm1.6$ &
5\rlap{$^g$}            &
22                      &
$-$1.54                 &
0.25                    &
0.60                    &
8.4                     \\
NGC\,7006               &
21 01 29.5              &
\llap{1}6 11 15         &
63.77                   &
$-$19.41                &
41.5                    &
0.05                    &
                        &
\llap{$-36$}$9.7\pm0.4$ &
4\rlap{$^g$}            &
17                      &
$-$1.63                 &
0.24                    &
0.38                    &
6.3                     \\
M\,15                   &
21 29 58.3              &
\llap{1}2 10 01         &
65.01                   &
$-$27.31                &
10.3                    &
0.10                    &
                        &
\llap{$-9$}$4.8\pm0.2$  &
\llap{1}3\rlap{$^{c,d}$} &
41                      &
$-$2.25                 &
0.07                    &
1.06                    &
\llap{2}1.5             \\
M\,2                    &
21 33 29.3              &
\llap{$-$}0 49 23       &
53.38                   &
$-$35.78                &
11.5                    &
0.06                    &
                        &
$5.1\pm2.0$             &
8\rlap{$^e$}            &
37                      &
$-$1.62                 &
0.34                    &
0.93                    &
\llap{2}1.5             \\
Pal\,13                 &
23 06 44.4              &
\llap{1}2 46 19         &
87.10                   &
$-$42.70                &
24.8\rlap{$^f$}         &
0.05                    &
                        &
\llap{3}$0.4\pm0.5$\rlap{$^b$} &
2\rlap{$^{a,b}$}        &
3                       &
$-$1.65                 &
0.48                    &
0.46                    &
2.2                     \\
\hline
\end{tabular}
\end{table*}

\section{Results}

We analysed the ``total'' ON$-$OFF spectrum and, where applicable, the
difference between the total ON spectrum and (only) the combined spectra taken
at the offset pointings in the sparse map. We shall refer to the latter as the
``mapping'' spectrum, whilst the ``pointing'' spectrum refers to the standard
ON/OFF position switching mode excluding the sparse map offset pointings.

\subsection{Measuring the ICM mass}

To convert measured brightness temperature into gas mass, one has to first
consider whether the emission is spatially resolved. To estimate the expected
radial extent of a cool ICM, we assume hydrostatic equilibrium:
\begin{equation}
\frac{{\rm d} p}{{\rm d} r} = - \rho_{\rm gas} \frac{G M}{r^2},
\end{equation}
where the gas has a pressure $p$ and density $\rho_{\rm gas}$, and the
gravitational potential is due to stellar mass $M$ enclosed by radius $r$.
Assuming an ideal, isothermal gas of temperature $T$, and using the Poisson
equation for the stars, we obtain:
\begin{equation}
\rho_\star = - \ \frac{k T}{4 \pi G m} \nabla^2 \ln \rho_{\rm gas},
\end{equation}
where the gas molecules have a mass $m$. We approximate the stellar mass
density, $\rho_\star$, by a singular isothermal sphere:
\begin{equation}
\rho_\star(r) = \rho_\star(r_0) \left(r_0/r\right)^2.
\end{equation}
Here $r_0$ is an arbitrary reference point for which we take the half-mass
radius, $r_{\rm M/2}$. By simple integration we obtain the density at that
point, $\rho_\star(r_{\rm M/2})$, as a function of half-mass radius and total
stellar mass, $M_\star$. Solving Eq.\ (2) for the gas density we obtain:
\begin{equation}
\rho_{\rm gas} \propto r^{-\alpha},
\end{equation}
where the exponent is given by:
\begin{equation}
\alpha = \frac{G m M_\star}{2 k T r_{\rm M/2}}.
\end{equation}
For typical values of $M_\star=10^5$ M$_\odot$, $r_{\rm M/2}=10^{17}$ m and
$m\sim2\times10^{-27}$ kg, the gas density profile follows exactly that of the
stars if $T\sim5,000$ K. Cooler gas will sink deeper into the core. We thus
conclude that the atomic or molecular ICM is not expected to be resolved as
the Arecibo beam is a few times larger than the half-mass radii of our target
clusters.

We thus emply the usual formula (e.g., Braun \& Burton 2000) to find the
neutral hydrogen mass per beam:
\begin{equation}
M_{\rm H} [{\rm M}_\odot] = 0.024\ d_\odot^2 \int T_{\rm B}(v)\
{\rm d}v,
\end{equation}
where the cluster distance, $d_\odot$, is in kpc, the brightness temperature,
$T_{\rm B}$, is in K and the velocities, $v$, are in km s$^{-1}$. We adopt a
main beam efficiency of 0.7 to convert antenna temperature units into a
brightness temperature scale (Heiles et al.\ 2001, with updates on
http://www.naic.edu/$\sim$phil/sysperf/sysperf.html).

The rms noise, $\sigma_{\rm T}$ (in antenna temperature), was computed in
50--100 km s$^{-1}$ spectral intervals at one or either side of the GC
velocity, excluding 20 km s$^{-1}$ centred at the GC velocity and avoiding
obvious signal. The noise can then be converted into the equivalent mass per
beam:
\begin{equation}
\sigma_{\rm M} [{\rm M}_\odot] = 0.034\ d_\odot^2\ \sigma_{\rm T}\
\left(\Delta v\ \sigma_{\rm rad}\right)^\frac{1}{2},
\end{equation}
where the channel width, $\Delta v$, and radial velocity dispersion,
$\sigma_{\rm rad}$, are in km s$^{-1}$. Noise levels are listed in Table 2.

%
%
\begin{table}
\caption[]{On-target integration times (pointing plus mapping) and rms noise
levels per beam. H\,{\sc i} rms values for M\,2 and Pal\,13 refer to the
difference spectrum with respect to the sparse mapping offset pointings
(Figs.\ 2 \& 3).}
\begin{tabular}{lccccc}
\hline\hline
                 &
                 &
\multicolumn{2}{c}{------ H\,{\sc i} ------} &
\multicolumn{2}{c}{------ OH ------} \\
Object           &
$t_{\rm int}$    &
$\sigma_{\rm T}$ &
$\sigma_{\rm M}$ &
$\sigma_{\rm T}$ &
$\sigma_{\rm F}$ \\
                 &
(min)            &
(mK)             &
(M$_\odot$)      &
(mK)             &
(mJy)            \\
\hline
NGC\,6934     &
58            &
5.5           &
0.1\rlap{5}   &
6.8           &
0.7           \\
NGC\,7006     &
48            &
6.6           &
0.9           &
7.5           &
0.8           \\
M\,15         &
\llap{1}00+30 &
3.3\rlap{$^\dagger$} &
0.0\rlap{5}   &
4.9           &
0.5           \\
M\,2          &
10+25         &
\llap{2}9.8   &
0.4\rlap{3}   &
9.7           &
1.0           \\
Pal\,13       &
25+20         &
\llap{1}0.1   &
0.3\rlap{4}   &
8.7           &
0.9           \\
\hline
\end{tabular}\\
$^\dagger$ rms = 1.4 mK in the smoothed $\times6$ spectrum of Fig.\ 1.
\end{table}

\subsection{Neutral hydrogen}

In Fig.\ 1 the H\,{\sc i} spectrum in the direction of the core of M\,15 is
displayed, together with a smoothed version obtained with a 6-channel running
boxcar. H\,{\sc i} emission is detected at precisely the velocity of M\,15.
The emission is resolved from the Galactic foreground emission, which only
interfers at $v_{\rm LSR}>-80$ km s$^{-1}$. The width of the emission is
similar to the radial velocity dispersion of M\,15, lending further support to
the reality of the detection. The integrated emission corresponds to a neutral
hydrogen mass of $M_{\rm H}\simeq0.3$ M$_\odot$, which is significant at a
level of more than 5 $\sigma$. Much less integration time was spent in mapping
mode than in pointing mode and the sparse map of M\,15 cannot therefore reveal
a smilarly weak signal as that shown in Fig.\ 1.

%
%
\begin{figure}
\centerline{\psfig{figure=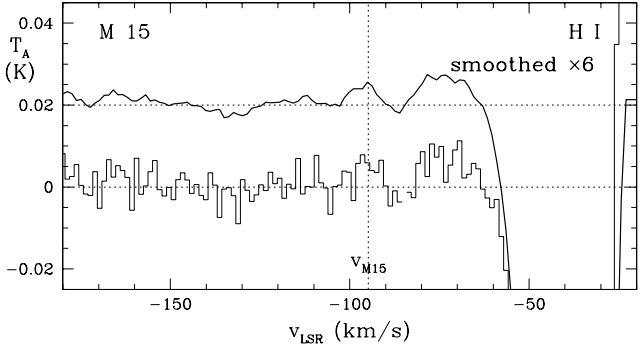,width=84mm}}
\caption[]{H\,{\sc i} 21 cm spectrum towards M\,15 (histogram). A smoothed
version of the spectrum is created by using a running boxcar of 6 channels,
and displayed with a vertical offset for clarity. With an original channel
width of 1.29 km s$^{-1}$ the smoothed spectrum has a velocity resolution of 8
km s$^{-1}$. A vertical line indicates the systemic velocity of M\,15, $v_{\rm
LSR}=-94.8$ km s$^{-1}$. Located outside of the Galactic foreground emission
(which is here negative after subtraction of brighter emission in the OFF
beam) is a faint peak of emission centred at the velocity of M\,15.}
\end{figure}

In Figs.\ 2 \& 3 are displayed the H\,{\sc i} spectra of M\,2 and Pal\,13,
together with the ``pointing'' OFF spectrum and the individual offset
pointings in the sparse map. If the same offset pointing was revisited then
the individual spectra were averaged. The middle panel shows the difference
between the ON spectrum and the combined OFF pointings from the sparse map,
along with the standard deviation ($\sigma$) amongst these OFF pointings. In
the bottom panel this ON$-$OFF spectrum is expressed in terms of $\sigma$ to
show the significance of excess emission in the core of the GC as compared to
the fluctuations in the surrounding emission.

For M\,2 the sparse mapping may have allowed us to detect H\,{\sc i} in its
core despite the strong Galactic foreground emission: the core has excess
emission over surrounding pointings at a velocity of $v_{\rm LSR}=+3$ km
s$^{-1}$ (Fig.\ 2), which is within the 2 km s$^{-1}$ uncertainty of the
velocity of M\,2. The emission reaches a nearly 4 $\sigma$ significance over
fluctuation in the surrounding emission. Another peak of excess emission at
$v_{\rm LSR}=+10$ km s$^{-1}$ does not appear to be significant as the same
emission component is seen in the S and NW offset pointings. Given the large
tidal radius of M\,2 it is still possible that this emission is extended and
physically associated with the cluster. The +3 km s$^{-1}$ component seems to
reside in the core of M\,2, and would correspond to a neutral hydrogen mass of
$M_{\rm H}\simeq3$ M$_\odot$.

%
%
\begin{figure}
\centerline{\psfig{figure=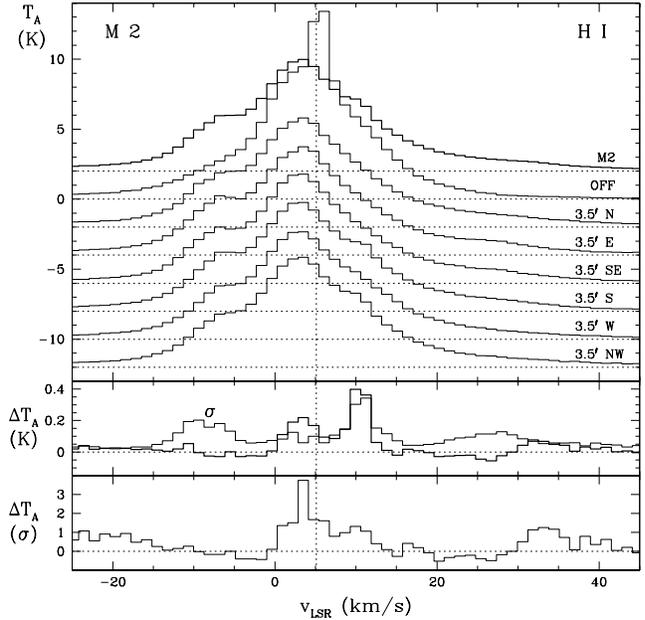,width=84mm}}
\caption[]{H\,{\sc i} 21 cm spectrum towards M\,2 (top panel, boldface). Below
that are displayed the position switching OFF spectrum and the six spectra
obtained at the individual OFF pointings in the sparse map. The spectra are
offset with respect to each other for clarity. In the middle panel are
displayed the M\,2 spectrum after subtracting the combined spectrum of the OFF
pointings in the sparse map (boldface) and the rms amongst the sparse mapping
OFF pointings ($\sigma$, thin line). Their ratio is displayed in the bottom
panel. A vertical line indicates the systemic velocity of M\,2, $v_{\rm
LSR}=5.1$ km s$^{-1}$. The channel width is 1.29 km s$^{-1}$. There is a hint
of emission at 3 km s$^{-1}$ centred at M\,2, and at 10 km s$^{-1}$ but
without showing a central concentration.}
\end{figure}

The radial velocity of Pal\,13 places it on the wing of strong Galactic
foreground emission. However, that emission varies very little across the
sparse map, and as a result there is a hint of extremely faint emission at
$v_{\rm LSR}=+33$ km s$^{-1}$ at a 2 $\sigma$ level and much stronger excess
emission at $v_{\rm LSR}\simeq+5$ km s$^{-1}$ (Fig.\ 3). Although the latter
reaches a significance of over 3 $\sigma$ with respect to the surrounding
emission fluctuations, this component suffers from much greater Galactic
contamination. Its large velocity difference of $\Delta v\simeq-25$ km
s$^{-1}$ casts doubt on a physical association with Pal\,13 as it exceeds the
escape velocity by almost an order of magnitude. Excess emission is, however,
seen at all velocities from $\sim$0 up to $\sim$40 km s$^{-1}$, and the two
``components'' only become separate when expressing their strengths in
$\sigma$.

%
%
\begin{figure}
\centerline{\psfig{figure=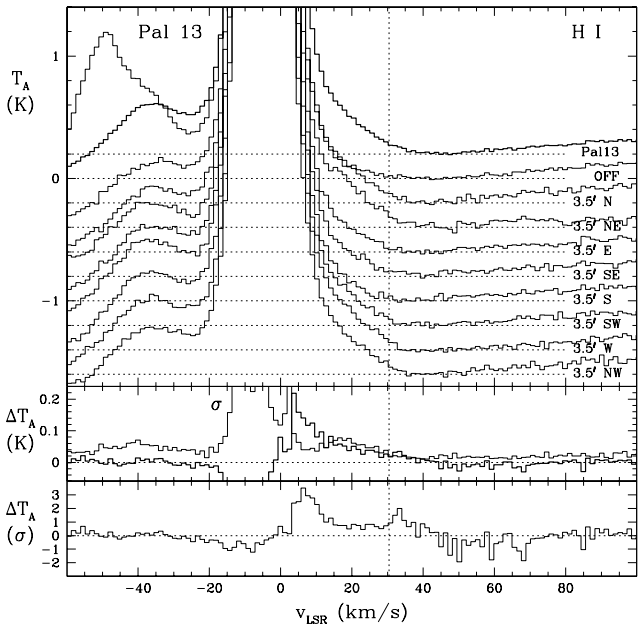,width=84mm}}
\caption[]{H\,{\sc i} 21 cm spectrum towards Pal\,13, displayed in a similar
way as Fig.\ 2, but now there are eight OFF pointings in the sparse map and
the systemic velocity of Pal\,13 is $v_{\rm LSR}=30.4$ km s$^{-1}$. There is a
hint of centrally concentrated emission at 5 and 33 km s$^{-1}$ and possibly
in between, but the Galactic foreground emission interfers strongly.}
\end{figure}

Emission from the Magellanic Stream behind Pal\,13 was detected at $v_{\rm
LSR}\simeq-382$ km s$^{-1}$ (see Appendix A), and an intermediate-velocity
cloud can be seen in the OFF beam at $v_{\rm LSR}\simeq-50$ km s$^{-1}$ (Fig.\
3).

No H\,{\sc i} was detected in NGC\,6934 or NGC\,7006, despite the very low
noise levels and large systemic velocities. The 5-$\sigma$ upper limit to the
neutral hydrogen mass in the Arecibo beam is $M_{\rm H}<0.8$ M$_\odot$ for
NGC\,6934, somewhat smaller than the upper limit obtained by Birkinshaw et
al.\ (1983). For NGC\,7006 the 5-$\sigma$ upper limit is $M_{\rm H}<5$
M$_\odot$. Given their small projected sizes it is unlikely that a significant
amount of neutral hydrogen mass was missed.

\subsection{Hydroxyl}

No OH emission was detected from any of the GCs in our sample, neither in the
main lines nor in the satellite lines. Dickey \& Malkan (1980) observed
NGC\,7006, M\,15 and Pal\,13 at 1665 and 1667 MHz at rms levels that
correspond to 6.6, 3.9 and 4.4 mJy for our velocity channel width: our search
is 5--8 times more sensitive (Table 2). Frail \& Beasley (1994) obtained upper
limits to the 1612 MHz emission at an equivalent rms level of $\sim$60 mJy.
Our search was more sensitive than this by two orders of magnitude.

\section{Discussion}

\subsection{The globular cluster ICM}

The deepest H\,{\sc i} survey to date was performed at Arecibo by Birkinshaw
et al.\ (1983). Not only is the new L-wide receiver more sensitive, but also
Birkinshaw et al.\ observed known radio sources at generally more than a
beam-width {\it away from the GC centre}. They will thus have missed gas
residing within the GC core. Pointing directly at the GC centres and
performing sparse mapping observations to remove Galactic foreground emission
greatly improved our chances of detecting the ICM.

We detect H\,{\sc i} emission in the core of M\,15. The excellent agreement
between the peak velocity and velocity dispersion of the H\,{\sc i} emission
and those of M\,15 strongly suggest that the neutral hydrogen in M\,15 is
associated with the ICM. It is unlikely to be associated with the PN Ps\,1 in
M\,15, as the velocity of Ps\,1 differs from that of M\,15 by $>20$ km
s$^{-1}$ (Rauch, Heber \& Werner 2002). The cold dust detected in M\,15 by
Evans et al.\ (2003) is almost certainly associated with the ICM (rather than
being of circumstellar nature) and the neutral hydrogen may be directly
associated with this dust component.

The H\,{\sc i} emission detected in at least one (M\,15) and possibly up to
three (M\,2 and Pal\,13) out of five GCs indicates that the ICM typically
contains $M_{\rm H}\lsim$1 M$_\odot$ of neutral hydrogen (of course, besides
any hydrogen there must be at least another 25 per cent in the form of
helium). This is roughly equivalent to the total amount of mass lost by a few
low-mass AGB stars, which happens on a timescale of $10^6$ yr. Stripping of
the ICM during Galactic plane crossing occurs on timescales that are two
orders of magnitude longer, and massive GCs such as M\,2 and M\,15 each
contain dozens of AGB stars. One would thus have expected to find more than
two orders of magnitude more ICM than what we detect here in neutral hydrogen.
This problem becomes even worse if we consider only 5-$\sigma$ upper limits to
our data, which correspond to $\sim1$ M$_\odot$.

Freire et al.\ (2001) argue that the stellar radiation field in a GC is
sufficient to ionize all hydrogen. They provide evidence for the presence of
$M_{\rm H}\simeq0.1$ M$_\odot$ in the form of plasma, in 47\,Tuc as well as in
M\,15. This is less than the mass of neutral hydrogen that we detect, and
therefore insufficient to solve the missing ICM mystery.

More hydrogen may be hidden in molecular form. The gas-to-dust mass ratio in
the circumstellar envelope of an AGB star is $\sim$200 at solar metallicity,
and scales approximately linearly with metallicity (van Loon et al.\ 2005).
The detection of $5\times10^{-4}$ M$_\odot$ of dust in M\,15 (Evans et al.\
2003), in combination with a gas-to-dust mass ratio of $\psi\sim$20,000 for
this metal-poor GC, would lead to a predicted $M_{\rm H}\simeq10$ M$_\odot$ of
gas, the majority of which must therefore be in molecular form. Such large
reservoirs of molecular gas are not supported by observations of CO, which
yield upper limits on the total gas mass of $M_{\rm gas}\sim$0.1 M$_\odot$.
The real gas-to-dust ratio is extremely uncertain, though. On the basis of the
detected amount of neutral hydrogen alone, one would estimate $\psi\sim$600
for the ICM of M\,15.

There are indications for Pal\,13 to be tidally disrupted (Siegel et al.\
2001), a process which may already have removed gas from its ICM. If the
H\,{\sc i} emission in our spectrum of Pal\,13 is real then its large velocity
spread of ${\Delta}v\sim$30 km s$^{-1}$ compared to its stellar velocity
dispersion and escape velocity of only 2--3 km s$^{-1}$ might imply that we
are witnessing the stripping of the gas from the GC. NGC\,6934 and NGC\,7006
are considerably more massive and tidal stripping at their current locations
in the Galactic halo is expected to be much weaker than for the low-mass GC
Pal\,13. Still, our upper limits to the neutral hydrogen content of these two
GCs are with a few M$_\odot$ tight enough to require some stripping mechanism
in addition to the passage through the Galactic plane. Especially NGC\,7006 is
quite distant from the galaxy and is therefore certain to have lived in
relative isolation for the past $10^8$ years.

Could a wind drive mass loss from the GC at a rate of $\dot{M}_{\rm
GC}\sim$10$^{-6}$ M$_\odot$ yr$^{-1}$? The wind speed in low-mass AGB stars is
too low to overcome the gravitational pull of a typical GC. However, the
combined stellar radiation field of the GC might be able to drive a wind for
much longer, pushing the gas out to a distance at which the local
gravitational acceleration has dropped sufficiently for the gas to become
gravitationally unbound. At an outflow velocity of 10 km s$^{-1}$ the gas
would have been dispersed over more than the Arecibo beamsize within a few
$10^5$ yr for a GC at a distance of 10 kpc. For this to work, the gas must
have sufficient opacity to be driven by radiation pressure. Although atomic
and molecular hydrogen are extremely inefficient for this, they may be
collisionally coupled with other sources of opacity such as dust grains or
ions. It would have been easier to explain an absence of dust and an abundance
of interstellar gas in this way as at low density the dust will be unable to
drag the gas with it, but at least in the case of M\,15 we do find ourselves
in that situation.

Isolated dwarf irregular (dIrr) galaxies are more gas-rich than dIrrs within
250 kpc of the Milky Way or M\,31, and the gas-less dwarf spheroidal galaxies
are almost exclusively found in the vicinity of either the Milky Way or M\,31.
This may be understood if ram pressure in a hot halo is effective in stripping
the dIrrs of their gas (Blitz \& Robishaw 2000). If true, then this could also
be a viable mechanism for stripping GCs of their gas as their orbits never
take them completely out of the galactic halo (Frank \& Gisler 1976; Lea \& De
Young 1976).

\subsection{Circumstellar masers in globular clusters}

The envelopes of mass-losing AGB stars provide an ideal environment for OH
masers to occur. The 18 cm transition can be pumped by the infrared radiation
arising from the circumstellar dust, and the slow cool wind ensures a long
velocity coherence path. The 1612 MHz line is normally the strongest maser in
well-developed outflows, but the 1665 MHz line can dominate at low mass-loss
rates when the OH is located closer to the warm stellar photosphere and the
1667 MHz line might be seen in circumstellar envelopes with short velocity
coherence paths. Thermal emission is orders of magnitude weaker than the maser
emission.

No maser emission was detected in any of the GCs that we looked at. Would we
have expected to detect any circumstellar masers? Assuming the same scaling
relation of the OH intensity with bolometric luminosity as found for the OH
masers detected in the Galactic centre and Large Magellanic Cloud (Marshall et
al.\ 2004), and assuming a bolometric magnitude of $-4<M_{\rm bol}<-3$ for a
low-mass thermal-pulsing AGB star, then the expected intensity of the OH maser
in a GC AGB star would be $\sim$0.1 Jy at the distance of M\,2 or M\,15. That
is more than an order of magnitude brighter than our 5-$\sigma$ upper limits.

The OH maser mechanism in the metal-poor outflows of low-mass GC AGB stars may
be less efficient due to the reduced infrared emission in a dust-poor wind.
The OH masers may also be fainter due to a lower abundance of hydroxyl as a
result of the lower abundance of oxygen and/or a higher photo-dissociation
rate in the stronger and less attenuated stellar radiation field in a GC.

\section{Summary}

We conducted the deepest survey to date, at H\,{\sc i} 21 cm and OH 1612,
1665, 1667 and 1720\,MHz with Arecibo, towards five Galactic globular
clusters. We present a tentative detection of 0.3 M$_\odot$ of neutral
hydrogen in the core of M\,15. This is a few orders of magnitude less than
what is expected to accumulate via stellar mass loss, and a removal mechanism
in addition to Galactic plane crossings is still required to explain the
paucity of the intra-cluster medium.

No OH maser emission was detected in any of the clusters, at a level which is
an order of magnitude below that expected from scaling known OH masers sources
to the likely properties of mass-losing AGB stars in globular clusters. The
faintness of globular cluster OH masers may be due to the low metallicity and
strong interstellar radiation field.

\section*{Acknowledgments}

We would like to thank the telescope operators at Arecibo Observatory for
their excellent support, Joana Oliveira for help with IDL and other matters,
and Carl Heiles for reading the manuscript. We also thank the referee for
her/his very helpful comments. Support by NSF grants AST-0097417 and
AST-9981308 is gratefully acknowledged. SS appreciates the hospitality of the
Keele Astrophysics Group during the observations.

\appendix

\section{The Northern tip of the Magellanic Stream behind Pal\,13}

H\,{\sc i} emission from the Magellanic Stream (Mathewson, Cleary \& Murray
1974) was detected behind Pal\,13 (Fig.\ A1). This section corresponds to the
northern tip of the stream, complex MS\,VI (Putman et al.\ 2003) and was
studied in detail in Stanimirovi\'{c} et al.\ (2002). To our knowledge, our
new observations represent the deepest H\,{\sc i} data of the Magellanic
Stream to date, comprising a total of 45 min on-source integration time and
reaching an rms noise level well below 0.01 K. The baseline was removed by
subtracting a smoothed version of the position-switching OFF spectrum, which
itself was completely featureless apart from the slow baseline structure. The
emission peaks at $v_{\rm LSR}\simeq-382$ km s$^{-1}$ and extends across
$\sim$70 km s$^{-1}$.

%
%
\begin{figure}
\centerline{\psfig{figure=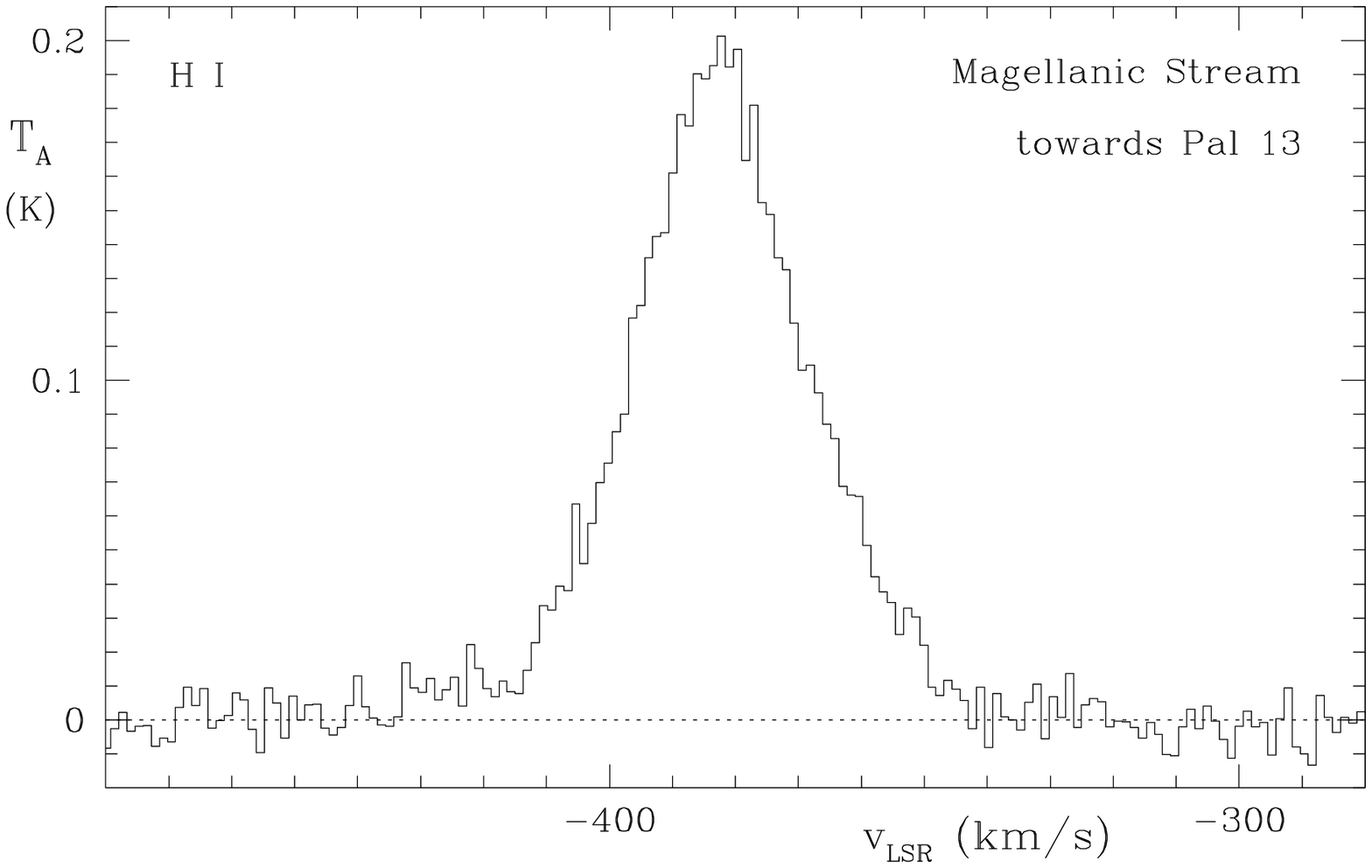,width=84mm}}
\caption[]{The Magellanic Stream detected in the H\,{\sc i} 21 cm line in the
direction of Pal\,13. The spectrum is the result of 45 min on-source
integration time, and a smoothed version of a featureless position-switching
OFF spectrum has been subtracted to remove the baseline. The channel width is
1.29 km s$^{-1}$.}
\end{figure}

\label{lastpage}

\end{document}